# IMPLEMENTATION OF THE DEPARTMENTAL AND PERIODICAL EXAMINATION ANALYZER SYSTEM

**Julius G. Garcia, Connie C. Aunario**

*Technological University of the Philippines, Manila, Philippines, julius.tim.garcia@gmail.com*

***Abstract** - Administering examinations both in public and private academic institutions can be tedious and unmanageable. The multiplicity of problems affecting the conduct of departmental and periodical examination can be greatly reduced by automating the examination process. The purpose of this action research is to provide an alternative technical solution in administering test through the use of Examination System. This software application can facilitate a plenitude of examinees for different subjects that implements a random questioning technique and can generate item analysis and test results. The Departmental and Periodical Examination System was developed using Visual Basic language. The software modules were tested using the functional testing method. Using the criteria and metrics of ISO 9126 software quality model, the system was evaluated by a group of students, teachers, school administrators and information technology professionals and has received an overall weighted mean of 4.56585 with an excellent descriptive rating. Therefore, the performance of the application software provides solution that can surmount the gargantuan problems of test administration and post-examination issues and performs all the operations specified in the objectives.*

## 1. Introduction

Primary and secondary schools conduct and administer periodical examinations for various courses on a quarterly basis or four terms in a school year. Likewise, major or departmental examinations in the tertiary level are conducted twice or thrice in a semester; prelim, midterm and finals respectively. The setting of examinations is common in different schools, colleges and universities around the world. Usually, the examinations are taken at the end of a course to determine the outcome for future advancement gauging the knowledge and skills acquired by the students. These involve the different types of testing and often objective tests; multiple choice questions (MCQ) items. Scoring in multiple choice questions is dichotomous and reliability and validity of items must be a concern.

Reliability defined (Berowitz et al., 2000) as the degree to which test scores for a group of test takers are consistent over repeated applications of a measurement procedure and hence are inferred to be dependable and repeatable for an individual test taker.

The performance data type and the metric in which the scores are express can affect the reliability of test scores (Yen & Candell, 1991). Increasing the length of examination can affect examinees due to fatigue and score becomes unreliable. Thus, it is important to know how student responded on each examination item, reflect the learning objectives and learning outcomes of a lesson, and be able to provide appropriate remediation.

However, this can be determined after examination has been administered. Moreover, the level of difficulty and discrimination of each item can be identified after the results were analyzed.

Therefore, it has been perceived to develop an examination analyzer system (EAS) that will provide quality examinations and learning as a whole.

### 1.1 Rationale

In order to achieve a focused grasp of the action research the following problems were answered:

1) What are the application features of the EAS?
2) What are the methods in formulating the examination items?
3) What are the processes involved in the examination item analysis?

## 1.2 Objective

The objective of the study is to:

1) Develop an EAS intended for departmental and periodical examination
2) Test the system using functional testing
3) Evaluate and implement the system.

## 1.3 Conceptual Framework

On the basis of the foregoing concepts, theories and findings related literature, studies and insights taken from them, a conceptual model was developed as illustrated in Figure 1.

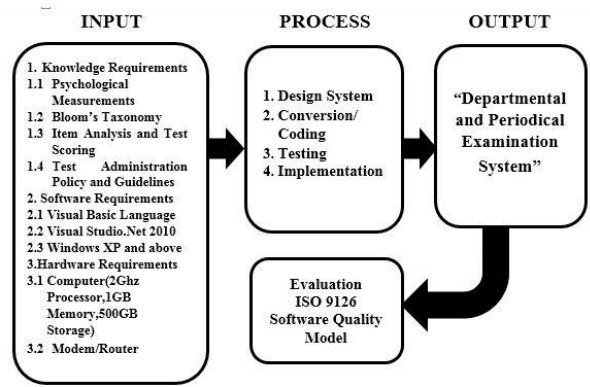

*Figure 1.* Conceptual Model of the Study

The conceptual model illustrates the different phases that the action research undergone through to realize its objectives.

The input phase constitutes the knowledge requirements, software requirements and hardware requirements. The knowledge in test measurements, construction and administration is critically significant in the development of the project.

In developing the system, the software requirements are as follows: (1) Visual Basic Language used for coding (2) Visual Studio. NET used as the integrated development environment (3) Window XP and above for platform integration.

The process phase includes the system design, system conversion and coding, testing and implementation. It explains how the design will be created as well as developing and enhancing the system. The system goes through the functional and system integration testing. The system is implemented in a LAN based network.

This results to the development of the Departmental and Periodical EAS.

## 2. Methodology

### 2.1 Participants

There were three groups of respondents. Group 1 consists of 30 undergraduate-student volunteers (23 females, 7 males), group 2 consists of 10 faculty members (8 females, 2 males) and group 3 consists of information technology experts (8 males, 2 females). All took the examination in the Departmental and Periodical EAS.

### 2.2 Resources and Materials

The examination was conducted at the computer laboratory with a total of 40 computer workstations. Fifty (50) multiple choice general knowledge questions were given to the participants to answer simultaneously in one hour.

Test scores were immediately acquired through the EAS right after the test. It provides individual test scores, group test results and item difficulty and discrimination results.

### 2.3 Procedure

At an initial meeting, participants were given informed consent. Each form contained an assigned examination identification number, password and workstation number. There are three groups consist of thirty (30) undergraduate-students, ten (10) faculty members and ten (10) information technology experts respectively. The first group which consists of undergraduate-students were assisted to the computer laboratory for the testing. The test administrator explained to the students the examination guidelines, rules and regulations.

Immediately after all participants finished answering the test, they were debriefed, requested to evaluate the system and dismissed outside of the computer laboratory.

The second and third group consists of faculty members, software engineers and system programmers were individually invited to take the examination and evaluate the system based on their time availability.

## 3. Findings

Based on the respective findings, the following data and information were unearthed and verified:

1) The EAS is deployed in a local area network and a local host server. It sets access level to the users through username and password encryption. It allows to create questions, set the time in minutes and the correct answer for each item and specify the passing rate. Each item can be categorized centered on the blooms taxonomy criteria (knowledge, comprehension, application, analysis, synthesis and evaluation) using the option buttons. The system can generate the: (1) examination identification number and password for examination access (2) individual and group test results per year level, section and subject (3) item results (number of students answered correct or incorrect for each item per blooms taxonomy criterion per year level, section and subject) (4) graphical interpretation of the test results per item, year level, section and subject.
2) In the existing system, the methods used in formulating and creating the examination items are multiple choice, true or false, matching type, short-written answer and essay.
3) The item analysis of examination items involves the following processes: (1) sorting of students name per year level per section per subject (2) tabulate number of item responses (3) computation of item difficulty and interpretation and computation of item discrimination and interpretation (4) generates and displays upper and lower group

## 4. Results and Discussion
### 4.1 System Test Results

The system has three main modules namely: the administrator module, faculty module and student module. Each module was tested and the results are shown on the Table 1.

*Table 1.* Summary of the Software Test Results

| Test Scenario | Actual Outcome | Remarks |
|---|---|---|
| Administrator Module | Access in all modules. Reset users password Add, Edit, Delete and Update users' account. | Passed |
| Faculty Module | Generate questions and set test component options View and sort test results per year level, section, and subject View item analysis result Print data results students profile | Passed |
| Student Module | Update and edit student information View his/her test result | Passed |

The testing was conducted by the proponent, a technical support and a quality assurance officer and has found the system functional.

### 4.2 Evaluation Results

The developed system was evaluated based on the ISO 9126 Software Quality Model. This aimed to assess the Functionality, Reliability, Usability, Efficiency, Maintainability and Portability of the system. The system was rated from 1 to 5 where 1 is the lowest and 5 is highest. The mean rating where then translated into qualitative interpretation using the following scale: 4.51 – 5.00 is Excellent; 3.51 – 4.50 is Very Good; 2.51 – 3.00 is Good; 1.51 – 2.50 Fair; and 1.00 – 1.50 is Poor.

The Table 2 shows the descriptive statistics of the software quality criteria. In terms of the functionality, the system received a total mean of 4.605 which implies an excellent descriptive rating from the three groups.

This further confirmed that the system is highly functional and has the ability to provide functions that meet the users need when it is under operation and execution. The graph shows the mean of functionality from the three groups with a mean of 4.6, 4.7 and 4.525 respectively as shown in Figure 2.

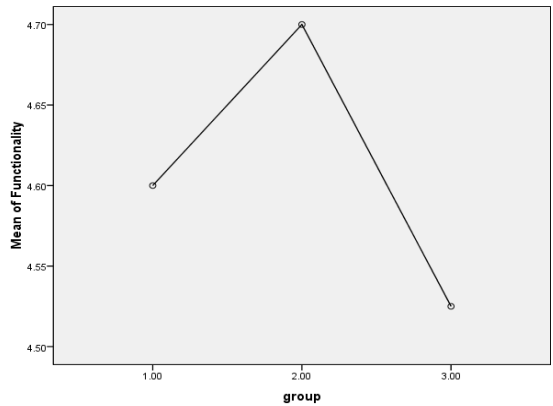

*Figure 2.* Mean of Functionality per Group

The system under specified conditions can maintain the specified level of performance due to the very low frequency of failure of the system. It can also withstand and recover from component, or environmental, failure.

Thus, the reliability of the system received a total mean of 4.6667 which implies a very good descriptive rating. It earned a mean of 4.6111, 4.8333 and 4.6667 from the three groups respectively as illustrated in the graph in figure 3.

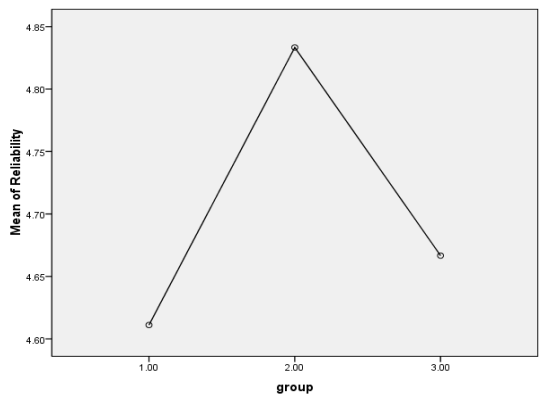

*Figure 3.* Mean of Reliability per Group

In terms of usability, the system demonstrated an excellent performance which earned a total mean 4.6467 and an excellent descriptive rating from the respondents. Each group scored the system with 4.6556, 4.7 and 4.5667 respectively as illustrated in Figure 4. Specified reasons are drawn: (1) there is an ease of the functions to be understood by the users (2) there is an ease of the software operation executed by the users (3) the software draws different users to learn it.

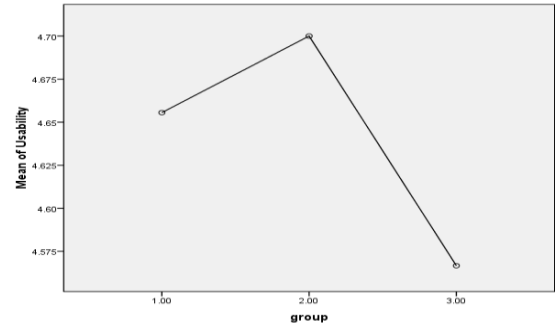

*Figure 4.* Mean of Usability per Group

Another factor considered was the efficiency characteristics of the system. Overall, the system's efficiency receives a total mean of 4.4867 which implies a very good descriptive rating which implies the software adheres and complies with the efficiency standards and conventions. The mean scores earned from the groups are 4.45, 4.65 and 4.3333 respectively as shown in Figure 5. It shows that the response time of the software for a given input is highly responsive and relatively fast. The resources used are always relative to the response of the software and vice-versa.

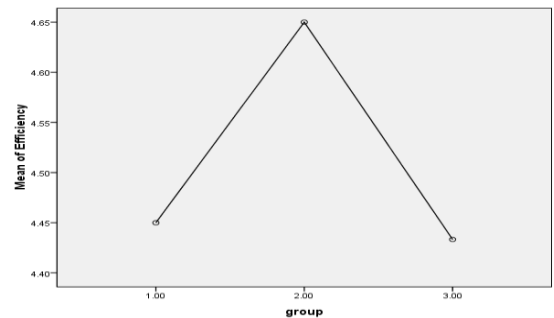

*Figure 5.* Mean of Efficiency per Group

In the case of the systems maintainability, it is easy to identify the root cause of failure within the software. In addition, the software is unlikely sensitive to any system changes. However, there is an effort needed to verify and test the system but not at all times. Generally, the system can subject itself to corrections, improvements or adaptation to changes in environment. Thus, the maintainability of the system earned a total mean of 4.4917 which implies a very good descriptive rating from the respondents. As illustrated in Figure 6, each group scored the system with a mean of 4.4917, 4.6, and 4.4 respectively.

Table 2. Descriptive Statistics of the Software Quality Criteria

| Software Criterion | | N | Mean | Std. Deviation | Std. Error | 95% Confidence Interval for Mean | | Minimum | Maximum |
|---|---|---|---|---|---|---|---|---|---|
| | | | | | | Lower Bound | Upper Bound | | |
| Functionality | 1 | 30 | 4.6 | 0.35111 | 0.0641 | 4.4689 | 4.7311 | 4 | 5 |
| | 2 | 10 | 4.7 | 0.2582 | 0.08165 | 4.5153 | 4.8847 | 4 | 5 |
| | 3 | 10 | 4.525 | 0.43221 | 0.13668 | 4.2158 | 4.8342 | 4 | 5 |
| | Total | | 4.605 | 0.35026 | 0.04953 | 4.5055 | 4.7045 | 4 | 5 |
| Reliability | 1 | 30 | 4.6111 | 0.35106 | 0.06409 | 4.48 | 4.7422 | 4 | 5 |
| | 2 | 10 | 4.8333 | 0.2357 | 0.07454 | 4.6647 | 5.0019 | 4.33 | 5 |
| | 3 | 10 | 4.6667 | 0.3849 | 0.12172 | 4.3913 | 4.942 | 4 | 5 |
| | Total | 50 | 4.6667 | 0.34339 | 0.04856 | 4.5691 | 4.7643 | 4 | 5 |
| Usability | 1 | 30 | 4.6556 | 0.33314 | 0.06082 | 4.5312 | 4.78 | 4 | 5 |
| | 2 | 10 | 4.7 | 0.33148 | 0.10482 | 4.4629 | 4.9371 | 4 | 5 |
| | 3 | 10 | 4.5667 | 0.31623 | 0.1 | 4.3405 | 4.7929 | 4 | 5 |
| | Total | 50 | 4.6467 | 0.32583 | 0.04608 | 4.5541 | 4.7393 | 4 | 5 |
| Efficiency | 1 | 30 | 4.45 | 0.3961 | 0.07232 | 4.3021 | 4.5979 | 4 | 5 |
| | 2 | 10 | 4.65 | 0.35746 | 0.11304 | 4.3943 | 4.9057 | 4 | 5 |
| | 3 | 10 | 4.4333 | 0.37843 | 0.11967 | 4.1626 | 4.704 | 4 | 5 |
| | Total | 50 | 4.4867 | 0.38662 | 0.05468 | 4.3768 | 4.5965 | 4 | 5 |
| Maintainability | 1 | 30 | 4.4917 | 0.39655 | 0.0724 | 4.3436 | 4.6397 | 4 | 5 |
| | 2 | 10 | 4.6 | 0.35746 | 0.11304 | 4.3443 | 4.8557 | 4 | 5 |
| | 3 | 10 | 4.4 | 0.4441 | 0.14044 | 4.0823 | 4.7177 | 4 | 5 |
| | Total | 50 | 4.495 | 0.39606 | 0.05601 | 4.3824 | 4.6076 | 4 | 5 |
| Portability | 1 | 30 | 4.5833 | 0.3675 | 0.0671 | 4.4461 | 4.7206 | 4 | 5 |
| | 2 | 10 | 4.625 | 0.13176 | 0.04167 | 4.5307 | 4.7193 | 4.5 | 4.75 |
| | 3 | 10 | 4.1000 | 0.39441 | 0.12472 | 3.8179 | 4.3821 | 3.5 | 4.75 |
| | Total | 50 | 4.4950 | 0.38956 | 0.05509 | 4.3843 | 4.6057 | 3.5 | 5 |

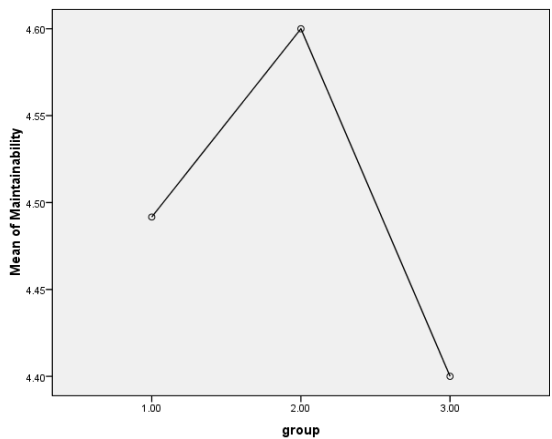

Figure 6. Mean of Maintainability per Group

In terms of Portability, the software can be transferred to another Windows OS platform only with such easiness and very less effort in software installation due to its size. It earned a total mean of 4.4950 which implies a very good descriptive rating. Figure 7 shows the graph of the mean of portability with a mean of 4.5833, 4.6250 and 4.100 received from each group respectively.

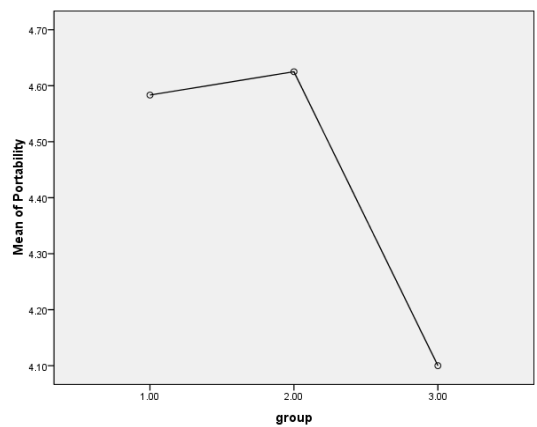

Figure 7. Mean of Portability ṭer Group

The overall mean for all the software criteria yielded an average rating of 4.56585 which clearly implies the performance of the system is excellent.

A one-way analysis of variance was conducted to determine significant difference between groups of respondents as illustrated on Table 3.

Table 3. ANOVA Results

| Software Criterion | | Sum of Squares | df | Mean Square | F | Sig. |
|---|---|---|---|---|---|---|
| Functionality | Between Groups | 0.155 | 2 | 0.077 | 0.622 | 0.541 |
| | Within Groups | 5.856 | 47 | 0.125 | | |
| | Total | 6.011 | 49 | | | |
| Reliability | Between Groups | 0.37 | 2 | 0.185 | 1.61 | 0.211 |
| | Within Groups | 5.407 | 47 | 0.115 | | |
| | Total | 5.778 | 49 | | | |
| Usability | Between Groups | 0.095 | 2 | 0.047 | 0.436 | 0.649 |
| | Within Groups | 5.107 | 47 | 0.109 | | |
| | Total | 5.202 | 49 | | | |
| Efficiency | Between Groups | 0.336 | 2 | 0.168 | 1.128 | 0.332 |
| | Within Groups | 6.989 | 47 | 0.149 | | |
| | Total | 7.324 | 49 | | | |
| Maintainability | Between Groups | 0.201 | 2 | 0.1 | 0.631 | 0.537 |
| | Within Groups | 7.485 | 47 | 0.159 | | |
| | Total | 7.686 | 49 | | | |
| Portability | Between Groups | 1.963 | 2 | 0.982 | 8.43 | 0.001 |
| | Within Groups | 5.473 | 47 | 0.116 | | |
| | Total | 7.436 | 49 | | | |

A one-way analysis of variance (ANOVA) showed no significant difference between groups and within groups at $p<0.05$ level $F(2,47)=0.622$, $p=0.541$ for the systems functionality; $F(2,47)=1.61$, $p=0.211$ for the systems reliability; $F(2,47)=0.436$, $p=0.649$ for the systems usability; $F(2,47)=1.128$, $p=0.332$ for the systems efficiency; $F(2,47)=0.631$, $p=0.537$ for the systems maintainability.

However, in the case of the systems portability, there is a significant difference between groups at $p<0.05$ level $F(2,47)=8.43$, $p=0.001$.

A Post-hoc test was conducted to determine which groups are significantly different. Tukey's result indicated significant difference between group 3 (information technology experts) and group 1 (undergraduate-students) at $p<0.05$ level $p=0.001$. It also indicated a significant difference between group 3 (information technology experts) and group 2 (faculty members) at $p<0.05$ level $p=0.003$. However, there was no significant difference between group 2 (faculty members) and group 1 (undergraduate-students) at $p<0.05$ level $p=0.940$. The significant difference in systems portability can be contributed to the technical expertise and knowledge of group 3 in evaluating the system in various platforms. However, the systems portability total mean of 4.4950 clearly indicates its very good performance using the intended platform.

5. Conclusion

The following conclusions are drawn in consideration with the objectives of the study and results of the evaluation.

1) The EAS can effectively facilitate users and automate the conduct of the departmental and periodical examination. It can be efficiently implemented in a local area network with such easiness. It runs in multiple workstations and randomly selects questions for each examinee to avoid cheating. Moreover, it can easily generate computed and graphical interpretation of course test results and item analysis results.

2) The test and evaluation results validated the systems performance.

3) The EAS demonstrated an excellent overall performance and can be effectively implemented for Departmental or Periodical examination both in the Secondary and Tertiary level.

**6. Recommendation**

1) Adequate training should be given to faculty members to familiarize and immerse them with the interface and functionality of the system.

2) It is also recommended to integrate EAS with the Academic Grading System❑